\documentclass[aps,pre,twocolumn]{revtex4-1} 	%looks like final product

\usepackage{graphicx, amsmath, amssymb, algorithmic, fancyvrb}
\usepackage{hyperref}
\hypersetup{ 
	colorlinks   = true
}

\def\be{\begin{equation}} 
\def\ee{\end{equation}} 
\newcommand \bea {\begin{eqnarray}} 
\newcommand \eea {\end{eqnarray}} 
\def \>{\rangle} 
\def \<{\langle}

\begin{document}
\title{The Community Simulator:\\ A Python package for microbial ecology}
\author{Robert Marsland III}
\affiliation{Department of Physics, Boston University, Boston, Massachusetts, USA}
\email{marsland@bu.edu}
\author{Wenping Cui}
\affiliation{Department of Physics, Boston College, Chestnut Hill, Massachusetts, USA}
\author{Joshua Goldford}
\affiliation{Bioinformatics Program, Boston University, Boston, Massachusetts, USA}
\author{Pankaj Mehta}
\affiliation{Department of Physics, Boston University, Boston, Massachusetts, USA}
%\email{pankajm@bu.edu}

\begin{abstract}
Natural microbial communities contain hundreds to thousands of interacting species. For this reason, computational simulations are playing an increasingly important role in microbial ecology. In this manuscript, we present a new open-source, freely available Python package called Community Simulator for simulating microbial population dynamics in a reproducible, transparent and scalable way. The Community Simulator includes five major elements: tools for preparing the initial states and environmental conditions for a set of samples, automatic generation of dynamical equations based on a dictionary of modeling assumptions, random parameter sampling with tunable levels of metabolic and taxonomic structure, parallel integration of the dynamical equations, and support for metacommunity dynamics with migration between samples. To significantly speed up simulations using Community Simulator, our Python package implements a new  Expectation-Maximization (EM) algorithm for finding equilibrium states of  community dynamics that exploits a recently discovered duality between ecological dynamics and convex optimization. We present data showing that this EM algorithm improves performance by between one and two orders compared to direct numerical integration of the corresponding ordinary differential equations. We conclude by listing several recent applications of the Community Simulator to problems in microbial ecology, and discussing possible extensions of the package for directly analyzing microbiome compositional data.
\end{abstract}

\maketitle

\section*{Background}

The last decade has seen a renewed interest in the study of microbial communities. Different environments can harbor diverse communities containing from hundreds to thousands of distinct microbes  \cite{EMP,HMP}. A central goal of community ecology is to understand the ecological processes that shape these diverse ecosystems. The diversity and function of ecosystems are affected by a wide variety of factors including energy and resource availability \cite{Loreau1995,Embree2015}, ecological processes such as competition between species\cite{Gause1935,MacArthur1970,Levin1970,Chesson1990} and stochastic colonization \cite{Chase2003, Jeraldo2012, Kessler2015, Vega2017}.

Microbial ecosystems also present several new challenges specific to microbes that are not usually addressed in the theoretical ecology literature. Classical models of community ecology (especially niche-based theories) have traditionally considered ecosystems with a few species and resources \cite{tilman_resource_1982, chesson2000mechanisms}. However,
microbial ecosystems often have thousands of species and hundreds of small molecules that can be consumed. It is unclear how the intuitions and results from these low-dimensional settings scale to  microbiomes.  It is known that diverse ecosystems can exhibit distinct emergent features and phase transitions not found in low-dimensional systems \cite{Fisher2014, Dickens2016, bunin2017ecological,Barbier2018}. Furthermore, classical ecological models usually assume a strict trophic layer separation, ignoring cross-feeding and syntrophy --  the consumption of metabolic byproducts of one species by another species. It is now becoming clear that cross-feeding is a central component of microbial ecosystems \cite{pacheco2019costless, Goldford2018, muscarella2019species,marsland2018available} and any ecological model must account for this phenomenon.

For these reasons, there is a need for new ways of understanding microbial ecosystems. One powerful approach for understanding complex systems is through simulations. However, simulating diverse microbial ecosystems presents some unique challenges. First, most ecosystems are mathematically represented by complicated coupled, non-linear ordinary differential equations. Simulating these systems in ecosystems with hundreds to thousands of species and metabolites becomes computationally difficult and time-consuming. Second, these dynamical models have thousands of parameters. One needs a principled and biologically realistic way of choosing such parameters. Third, explaining real data requires incorporating ecological processes such as stochastic colonization that play an important role in shaping community structure and dynamics. Finally, we need to be able to incorporate spatial and population level structures in an experimentally realistic way. 

\begin{figure*}[t]
	\includegraphics[width=17cm]{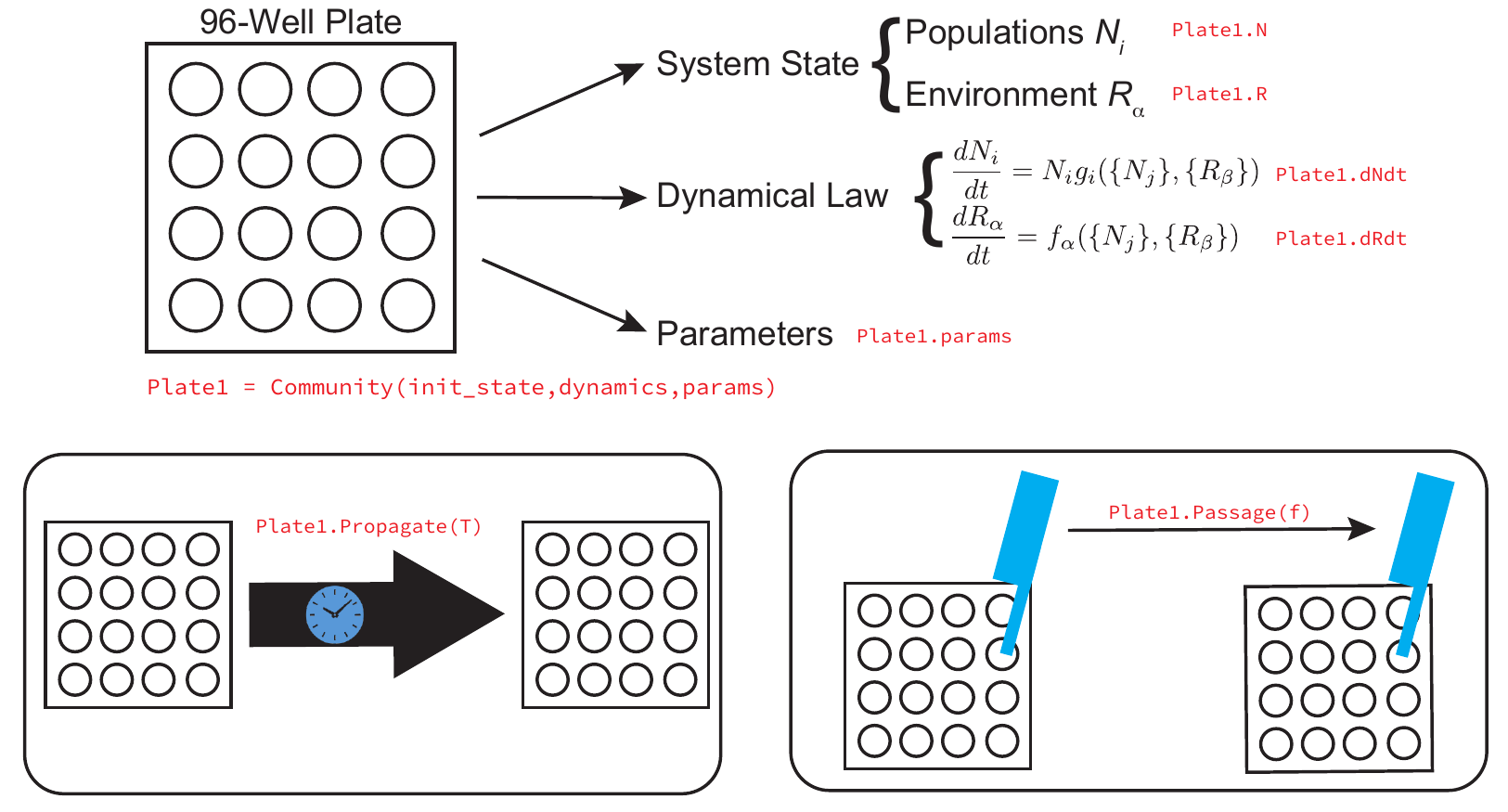}
	\caption{{\bf The  five elements of the Community Simulator.}
		The core object of the Community Simulator is a virtual $n$-well plate, holding $n$ independent well-mixed microbial communities. This plate has three properties: its current state, a dynamical law for the population dynamics, and a set of parameters. Once a plate is initialized, two actions can be performed on it: propagation in time using the given dynamical law, and passaging of given fractions of the contents of each well to fresh wells on a replacement plate. For some models, the equilibrium state of the population dynamics can also be found directly using a new algorithm summarized in Fig~\ref{EM} below.}
	\label{fig1}
\end{figure*}

Recently, we presented a powerful minimal model of microbial ecosystems that addresses these concerns \cite{Goldford2018, marsland2018available}. Furthermore, we have found a mathematical mapping between ecological dynamics and constrained optimization that can be used to accelerate simulations of many large ecosystems \cite{mehta2018constrained,Marsland2019a}. In this paper, we present a new open-source Python package for microbial ecology called Community Simulator that implements these theoretical advances, making it easy to simulate complex microbial communities in a variety of experimentally relevant settings.

\section*{Implementation}

The architecture of Community Simulator is inspired by the parallel experiments commonly performed with 96-well plates, as illustrated in Fig~\ref{fig1}. The central object of the package is a \verb+Community+ class, whose instances are initialized by specifying the initial population sizes and resource concentrations for each parallel ``well,'' along with the functions and parameters that define the population dynamics. The initial state, dynamical equations and parameters can all be generated automatically from a dictionary of modeling assumptions, or custom-built by the user. Each instance of this class represents an $n$-well plate, containing $n$ well-mixed, non-interacting communities. Once initialized, the state of the plate can be updated in one of two ways. \verb+Propagate(T)+ propagates the system for a time $T$ by integrating the supplied dynamical equations, and \verb+Passage(f)+ builds a replacement plate by adding a fraction $f_{\mu\nu}$ of the contents of each old well $\nu$ to each new well $\mu$. In the final section below, we will discuss a third method \verb+SteadyState()+, which can find the fixed point of the dynamics in some models without numerical integration.

The package also includes some functions for analysis of the simulation results, including a variety of measures of alpha diversity, as well as extraction of energy flux networks, effective interaction coefficients, and sensitivities to parameter perturbations. 

In the following sections, we describe the functionality of each element of the package in turn. We particularly focus on the tools for generating the dynamical equations and the parameter sets, explaining how increasing levels of biological realism can be progressively incorporated. Each section has a corresponding segment in the Jupyter notebook \verb|Tutorial.ipynb| included with the Community Simulator package. This notebook contains all the code and parameters for generating the figures found in the paper.

\begin{figure*}[t]
	\includegraphics[width=17cm]{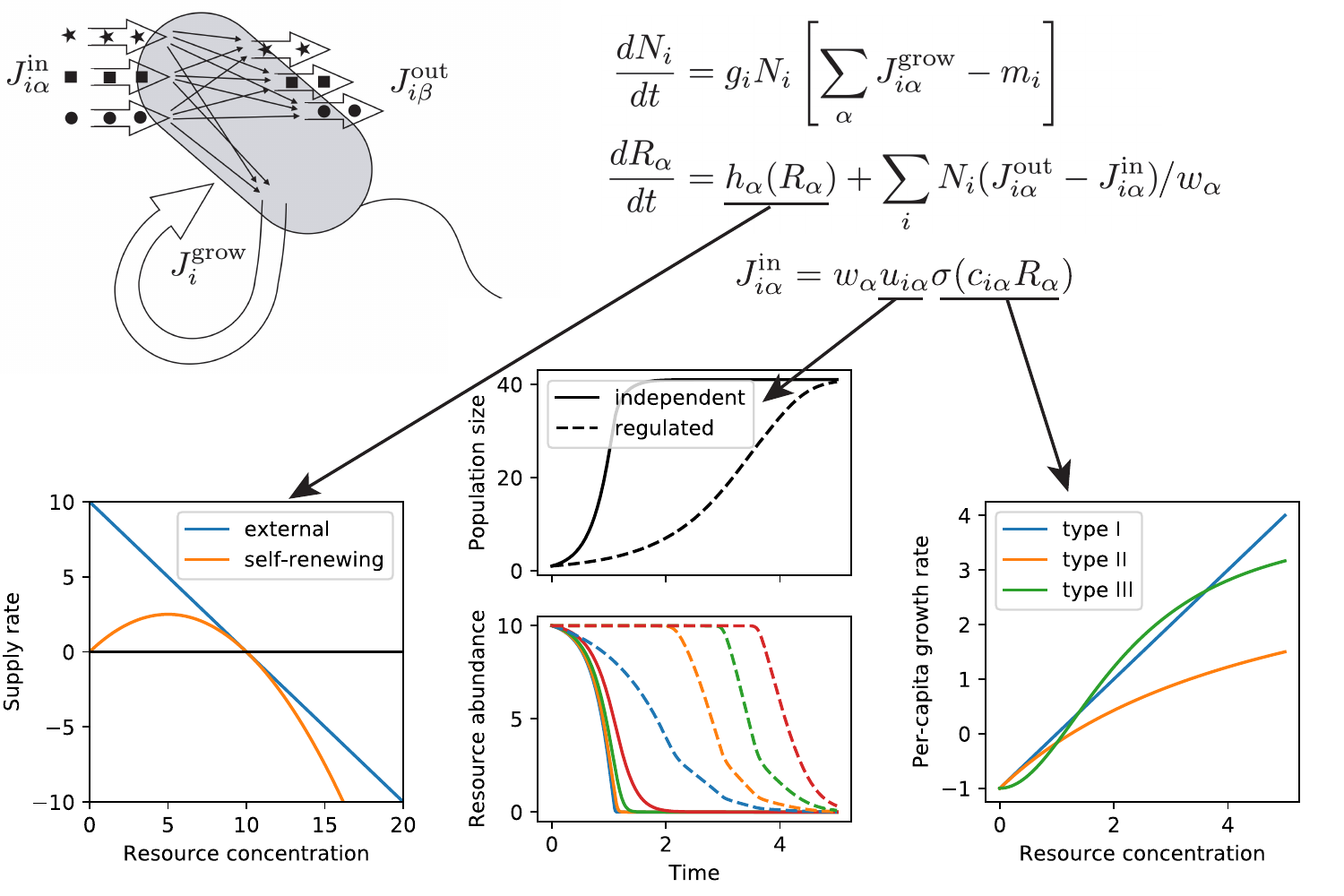}
	\caption{{\bf Constructing the dynamical law.} The MicroCRM models the growth and metabolism of $S$ microbial species in terms of energy fluxes $J_{i\alpha}^{\rm in}, J_{i\beta}^{\rm out}, J_i^{\rm grow}$, mediated by import, export and chemical transformation of $M$ substitutable resources. Specification of the resource dynamics and of the dependence of import rates on the resource concentrations requires three additional modeling choices, represented by the three arrows. First, the intrinsic dynamics of the resources can either be a linear model of a fixed external input flux and dilution rate, or a logistic model of self-renewing resources, which was employed in MacArthur's original CRM. The left-hand plot shows the supply rate as a function of resource concentration for these two options. Second, the import rates from the different resource types can be independent, or globally regulated in such a way as to preferentially consume the resource that is currently most abundant. The middle plot shows timeseries of consumer and resource abundances in the presence and absence of regulation, with all other parameters held fixed. Third, the dependence of import rates on resource concentration can take a linear (Type-I), Monod (Type-II) or Hill (Type-III) form. The right-hand plot shows the growth rate as a function of resource concentration for these three choices.}
	\label{dynamics}
\end{figure*}

\subsection*{Constructing the initial state}
The state of a \verb+Community+ instance is contained in a pair of Pandas data frames (\url{https://pandas.pydata.org/}, \cite{mckinney-proc-scipy-2010}), one of size $S_{\rm tot} \times n$ for the microbial population sizes $N_i$ ($i = 1, 2, \dots S_{\rm tot}$) and one of size $M \times n$ for the resource abundances $R_\alpha$ ($\alpha = 1, 2, \dots M$). Each row of the data frame corresponds to a different species or resource type, while each column corresponds to a different well.

\begin{table*}[t]
	\begin{tabular}{|c|l|}
		\hline
		$N_i$ & population density of species $i$ (individuals/volume)\\
		\hline
		$R_\alpha$ & Concentration of resource $\alpha$ (mass/volume)\\
		\hline
		$c_{i\alpha}$ & Uptake rate per unit concentration of resource $\alpha$ by species $i$ (volume/time)\\
		\hline
		$D_{\alpha\beta}$ & Fraction of byproducts from resource $\beta$ converted to $\alpha$ (unitless)\\
		\hline
		$g_i$ & Conversion factor from energy uptake to growth rate (1/energy)\\
		\hline
		$w_\alpha$ & Energy content of resource $\alpha$ (energy/mass)\\
		\hline
		$l_\alpha$ & Leakage fraction for resource $\alpha$ (unitless)\\
		\hline
		$m_i$ & Minimal energy uptake for maintenance of species $i$ (energy/time)\\
		\hline
		$R_\alpha^0$ & Intrinsic equilibrium abundance of resource $\alpha$ (mass/volume)\\
		\hline
		$\tau_\alpha$ & Timescale for externally supplied resource turnover (time)\\
		\hline
		$r_\alpha$ & Rate of resource self-renewal (volume/mass/time)\\
		\hline
		$n$ & Hill coefficient for functional response (unitless) \\
		\hline
		$\sigma_{\rm max}$ & Maximum input flux (mass/time)\\
		\hline 
		$n_{\rm reg}$ & Hill coefficient for metabolic regulation (unitless)\\
		\hline
	\end{tabular}
	\caption{\bf Parameters and units for the Microbial Consumer Resource Model.}
	\label{tab:param}
\end{table*}

The function \verb+MakeInitialState+ automatically creates data frames \verb+N0,R0+ of initial population sizes and resource abundances corresponding to some common experimental scenarios, specified in a dictionary of assumptions. The initial species abundances are supplied by a stochastic process that is agnostic to species identity. This roughly captures the various dispersal mechanisms including mechanical disturbances and turbulent flow that convey microbial cells to new environments. Specifically, random subsets of $S$ species from the regional pool of size $S_{\rm tot}$ are supplied to the $n$ wells of the plate. The population sizes of these species are set to 1 by default, and can be rescaled afterwards if desired.

%\begin{figure*}[!h]
%	\includegraphics[width=17cm]{init_state.pdf}
%	\caption{{\bf Preparing the initial state.} Stochastic colonization, food sources, dataframes.}
%	\label{init_state}
%\end{figure*}

Initial resource abundances are generated by \verb+MakeInitialState+ based on a Biolog plate scenario, where each well is supplied with a single carbon source. The assumptions dictionary specifies the identity and quantity of the carbon source for each well. Arbitrarily chosen resource abundances can of course be directly supplied to the \verb|Community| instance instead, to simulate more general conditions. 

To capture coarse-grained metabolic structure, the $M$ resources can be assigned to $T$ classes (e.g. sugars, amino acids, etc.), each with $M_A$ resources where $A=1, \ldots T$ and $\sum_A M_A=M$. These class labels become functionally relevant by biasing the sampling of consumption preferences and byproduct stoichiometry, as will be described below. Likewise the $S_{\rm tot}$ species can be assigned to $F$ families, with $F\leq T$, and each family preferentially consuming resources from a different resource class. A generalist family can also be included, with $S_{\rm gen}$ species and no preferred resource class, so that $S_{\rm gen} + \sum_A S_A = S_{\rm tot}$.

\subsection*{Generating the dynamical equations}
Instances of the \verb|Community| class can be initialized with any set of differential equations, which are specified as functions of the system state that return the time derivatives $dN_i/dt$ and $dR_\alpha/dt$. The package includes tools for constructing these functions automatically based on a dictionary of assumptions. These built-in dynamics are based on the recently introduced Microbial Consumer Resource Model (MicroCRM) \cite{Goldford2018,marsland2018available} illustrated in Fig~\ref{dynamics}, which generalizes the classic consumer resource model of MacArthur and Levins \cite{MacArthur1970} to the microbial context by allowing organisms to release metabolic byproducts. Table \ref{tab:param} lists all the parameters of this family of models, along with the corresponding units.

In order to provide a general-purpose set of models that produce physically reasonable results, the MicroCRM assumes that all resource types are substitutable, and can all be converted to a common energy currency. This allows us to enforce energy conservation, preventing communities from bootstrapping themselves to large population sizes using metabolic secretions with no external resource supply. It also eliminates the need to specify in detail how each resource type interacts with all the others within the consumer metabolism. If such interactions are important for capturing a given experimental phenomenon, the built-in dynamics cannot be used, and custom functions must be written for $dN_i/dt$ and $dR_\alpha/dt$. The tutorial notebook included with the package contains an example of this kind, using Liebig's Law of the Minimum to model phytoplankton dynamics.

\subsubsection*{Energy fluxes and growth rates}
We begin by defining an energy flux into a cell $J^{\mathrm{in}}$, an energy flux that is used for growth $J^{\mathrm{growth}}$, and an outgoing energy flux due to byproduct secretion $J^{\mathrm{out}}$. Energy conservation requires
\be
\label{eq:energy}
J^{\mathrm{in}}=J^{\mathrm{growth}}+J^{\mathrm{out}}
\ee
for any reasonable metabolic model. It is useful to denote the input and output energy fluxes that are consumed/secreted in metabolite $\beta$ by $J_{\beta}^{\mathrm{in}}$ and $J_{\beta}^{\mathrm{out}}$ respectively.
We can define corresponding mass fluxes by 
\be
\nu_{\beta}^{\mathrm{out}}\equiv J_{\beta}^{\mathrm{out}}/w_\beta
\ee
and
\be
\nu_{\beta}^{\mathrm{in}}\equiv J_{\beta}^{\mathrm{in}}/w_\beta
\ee
where the conversion factor $w_\beta$ measures the energy density of metabolite $\beta$. In general, all these fluxes depend on the consumer species under consideration, and will carry an extra Roman index $i$ indicating the species.

We assume that a fixed quantity $m_i$ of power per cell is required for maintenance of species $i$, and that the per-capita growth rate is proportional to the remaining energy flux $(J^{\rm growth}-m_i)$, with proportionality constant $g_i$. Under these assumptions, the time-evolution of the population size $N_i$ of species $i$ can be modeled using the equation
\be
{dN_i \over dt}= g_i N_i (J_i^{\mathrm{growth}} -m_i).
\ee
We can model the resource dynamics by functions of the form
\be
{dR_\alpha \over dt} =h_\alpha (R_\alpha) - \sum_j N_j \nu_{j \alpha}^{\mathrm{in}} + \sum_j N_j \nu_{j \alpha}^{\mathrm{out}},
\ee
where the function $h_\alpha$ describes the resource dynamics in the absence of consumers. The Community Simulator has two kinds of default resource dynamics: externally supplied and self-renewing. For externally supplied resources, we take a linearized form of the dynamics:
\be
h_\alpha^{\mathrm{external}} (R_\alpha) = \tau_\alpha^{-1}(R_\alpha^0 - R_\alpha)
\ee
while for self-renewing we take a logistic form
\be
h_\alpha^{\mathrm{self-renewing}} (R_\alpha) = r_\alpha R_\alpha (R_\alpha^0-R_\alpha).
\ee
Finally, the intrinsic resource dynamics can also be turned off, with $h_\alpha^{\rm off} = 0$, to simulate resource depletion in a closed community with no resupply.

\SaveVerb{make}|MakeMatrices|
\begin{table*}[t]
	\begin{tabular}{|c|l|}
		\hline
		$M$ & Number of resources\\
		\hline
		$T$ & Number of resource classes\\
		\hline
		$S_{\rm tot}$ & Number of microbial species in regional pool\\
		\hline
		$F$ & Number of specialist families\\
		\hline
		$S$ & Number of microbial species initially present in each local community\\
		\hline
		$\mu_c$ & Mean sum over a row of the preference matrix $c_{i\alpha}$\\
		\hline
		$\sigma_c$ & Standard deviation of sum over a row for Gaussian or Gamma $c_{i\alpha}$\\
		\hline
		$c_0$ & Low consumption level for Binary $c_{i\alpha}$\\
		\hline
		$c_1$ & High consumption level for Binary $c_{i\alpha}$\\
		\hline
		$q$ & Fraction of consumption capacity allocated to preferred resource class\\
		\hline
		$s$ & Sparsity of metabolic matrix\\
		\hline
		$f_w$ & Fraction of secreted byproducts allocated to ``waste'' resource class\\
		\hline
		$f_s$ & Fraction of secreted byproducts allocated to same resource class\\
		\hline
	\end{tabular}
	\caption{{\bf Definitions of global parameters used for constructing random ecosystems.} Values of these parameters are supplied as a Python dictionary to the function \protect\UseVerb{make}, which generates randomly sampled consumer preference and metabolic matrices.}
	\label{tab:meta}
\end{table*}

\begin{figure*}[t]
	\includegraphics[width=17cm]{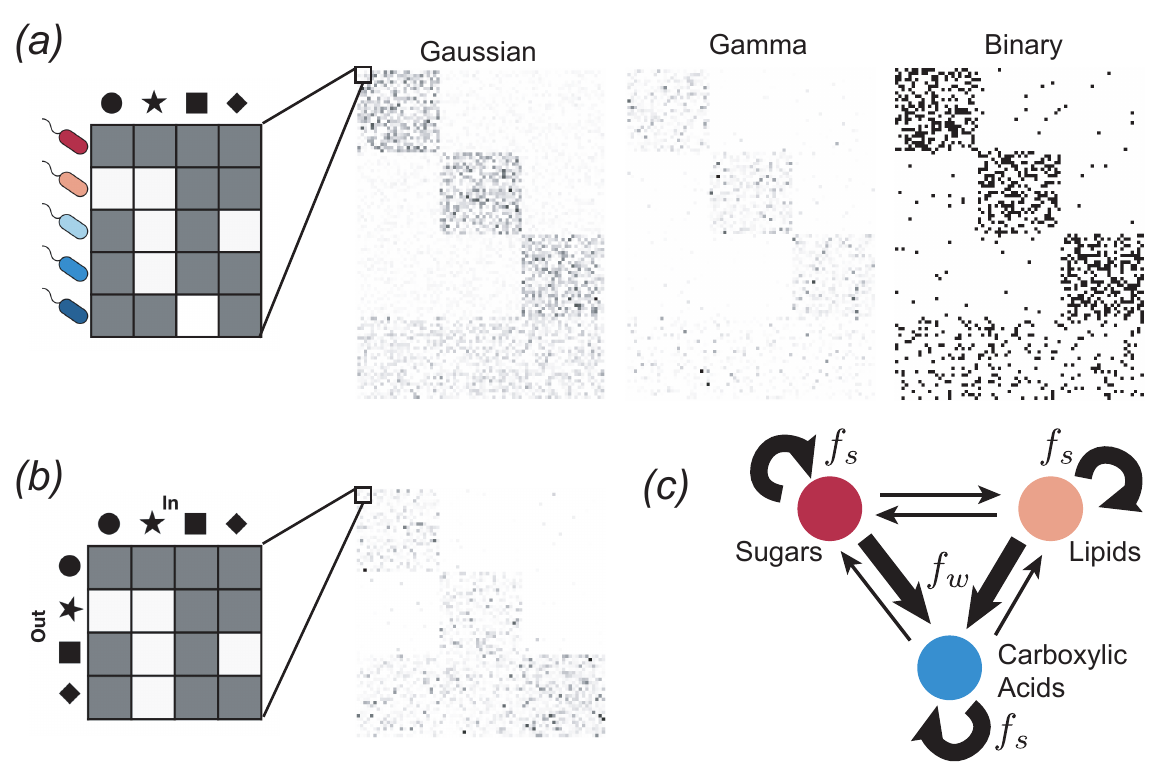}
	\caption{{\bf Sampling parameters and adding metabolic structure.} \emph{(a)} Sampling the consumer preference matrix $c_{i\alpha}$. Each row corresponds to a different microbial species, and the value of each entry in the row specifies the preference level of that species for a given resource. An example of each of the three sampling choices is shown, with white pixels representing $c_{i\alpha} = 0$ and darker pixes representing larger values. The examples have $F = 3$ consumer families with specialism level $q=0.9$, each with $S_A = 25$ species, plus a generalist family with $S_{\rm gen} = 25$ species. \emph{(b)} Sampling the metabolic matrix $D_{\alpha\beta}$. Each column represents the allocation of output fluxes resulting from metabolism of a given input resource. This example has $T = 3$ resource classes, and an effective sparsity $s = 0.05$. \emph{(c)} Diagram of three-tiered metabolic structure. A fraction $f_s$ of the output flux is allocated to resources from the same resource class as the input, while a fraction $f_w$ is allocated to the ``waste'' class (e.g., carboxylic acids). In the example of the previous panel, allocation fractions were $f_s=f_w=0.49$. }
	\label{params}
\end{figure*}

\subsubsection*{Input fluxes and output partitioning}
We now specify the form of the input fluxes $\nu_{\beta}^{\mathrm{in}}$, and of the relationships among input, output and growth that define the metabolism. We start by assuming that all resource utilization pathways are independent, resulting in input fluxes of the form
\be
\nu_{i\alpha}^{\mathrm{in}} =  \sigma(c_{i\alpha}R_\alpha)
\ee
where $\sigma$ is a single-valued function encoding the relationship between resource availability and uptake rates. The community simulator implements three kinds of response functions: Type-I, linear response functions where 
\be
\sigma_I(x)=x,
\ee
a Type-II saturating Monod function,
\be
\sigma_{II}(x) = {x \over 1+{x \over \sigma_{\rm max}}}
\ee
and a Type-III Hill or sigmoid-like function
\be
\sigma_{III}(x) = {x^n \over 1 +{x^n \over  \sigma_{\rm max}}},
\ee
where $n>1$.

To obtain the output fluxes, we define a leakage fraction $l$ such that
\be
J^{\mathrm{out}}= l J^{\mathrm{in}}.
\ee
We allow different resources to have different leakage fractions $l_\alpha$. A direct consequence of energy conservation (Eq (\ref{eq:energy})) is that
\be
J_i^{\mathrm{growth}}= \sum_\alpha (1-l_\alpha) J_{i \alpha}^{\mathrm{in}} = \sum_\alpha (1-l_\alpha) w_\alpha  \sigma(c_{i\alpha}R_\alpha)
\ee
Finally, we denote by $D_{\beta \alpha}$ the fraction of the output energy that is contained in metabolite $\beta$ when a cell consumes $\alpha$. Note that by definition $\sum_{\beta} D_{\beta\alpha} = 1$. The total energy output in metabolite $\beta$ is thus 
\be
J_{i\beta}^{\mathrm{out}} = \sum_{\alpha} D_{\beta \alpha} l_\alpha J_{i\alpha}^{\mathrm{in}}= \sum_{\alpha} D_{\beta \alpha} l_\alpha w_\alpha \sigma(c_{i \alpha} R_\alpha).
\ee
This also yields
\be
\nu_{i\beta}^{\mathrm{out}} =  \sum_{\alpha} D_{\beta \alpha} l_\alpha  {w_\alpha \over w_\beta} \sigma(c_{i \alpha} R_\alpha)
\ee
We are now in position to write down the full dynamics in terms of these quantities:
\bea
{dN_i \over dt}&=& g_i N_i \left[\sum_\alpha  (1-l_\alpha) w_\alpha \sigma(c_{i \alpha} R_\alpha) -m_i\right] \nonumber\\
{dR_\alpha \over dt} &=& h_\alpha (R_\alpha) - \sum_j N_j \sigma(c_{j \alpha} R_\alpha)\nonumber\\
&&+  \sum_{j \beta} N_j   \sigma(c_{j \beta} R_\beta) \left[ D_{\alpha \beta}{w_\beta \over w_\alpha} l_\beta \right]
\label{eq:dynamics}
\eea
Notice that when $\sigma$ is Type-I (linear) and $l_\alpha =0$ for all $\alpha$ (no leakage or byproducts), this reduces to MacArthur's original model  \cite{MacArthur1970}.

\subsubsection*{Metabolic regulation}
The package can also generate dynamics for active metabolic regulation, which allocates a higher fraction of import capacity to nutrients with higher available energy flux. This regulation is implemented through a series of weight functions for resource $\alpha$ that reflect how much of the utilizable energy in the environment is in resource $\alpha$
\be
u_{i \alpha}^{\mathrm{in}-w}= {(w_\alpha c_{i \alpha}R_\alpha)^{n_{\rm reg}} \over \sum_{\beta}  (w_\beta c_{i \beta}R_\beta)^{n_{\rm reg}} },
\ee
with $n_{\rm reg}$ a Hill coefficient that tunes steepness.
Another option is to regulate based on the fraction of biomass contained in resource $\alpha$,
\be
u_{i \alpha}^{\mathrm{in}-\nu}= {( c_{i \alpha}R_\alpha)^{n_{\rm reg}} \over \sum_{\beta}  (c_{i \beta}R_\beta)^{n_{\rm reg}} }
\ee
For the metabolically regulated model,  we define the input fluxes by
\be
\nu_{\beta}^{\mathrm{in}} = u_{i \beta}^\mathrm{in} \sigma(c_{i \beta} R_\beta)
\ee
Then, we can follow the exact same procedure as above. This yields the equations
\bea
{dN_i \over dt}&=& g_i N_i \left[\sum_\alpha   (1-l_\alpha) w_\alpha u_{i \alpha}^\mathrm{in}  \sigma(c_{i \alpha} R_\alpha) -m_i\right] \nonumber\\
{dR_\alpha \over dt} &=& h_\alpha (R_\alpha) - \sum_j N_j u_{j \alpha}^\mathrm{in}  \sigma(c_{j \alpha} R_\alpha)  \nonumber\\
&&+ \sum_{j \beta} N_j  u_{j \beta}^\mathrm{in} \sigma(c_{j \beta} R_\beta) \left[ l_\beta D_{\alpha \beta}{w_\beta \over w_\alpha} \right]
\eea
These equations are generated by the functions \verb+MakeConsumerDynamics+ and \verb+MakeResourceDyanamics+, based on the user's specification of the resource replenishment mode $h$, the response function $\sigma$, and the regulation mode $u$. 

\subsection*{Sampling the parameters}
The MicroCRM contains a large number of parameters: the $S_{\rm tot}$-dimensional vectors $g_i$ and $m_i$, the $M$-dimensional vectors $R_\alpha^0, l_\alpha, w_\alpha$ and $\tau_\alpha$ or $r_\alpha$, the $S_{\rm tot}\times M$ consumer preference matrix $c_{i\alpha}$ and the $M\times M$ metabolic matrix $D_{\alpha\beta}$. Some modeling choices require a small number of additional parameters: the maximal uptake rate $\sigma_{\rm max}$ for Type-II and Type-III growth, and the exponents $n$ for Type-III growth and $n_{\rm reg}$ for metabolic regulation. A dictionary containing all these parameters must be supplied to the \verb+Community+ instance upon initialization. A list of dictionaries may be supplied instead, to allow different wells to have different parameters.

The package contains a function \verb+MakeMatrices+ for generating the two matrices, which contain most of the ecological structure, based on a dictionary of modeling assumptions summarized in Table \ref{tab:meta}. The output of this function is illustrated in Fig~\ref{params} and described in detail below. 

\SaveVerb{pass}|Passage|
\SaveVerb{prop}|Propagate|
\SaveVerb{com}|Community|
\SaveVerb{tut}|Tutorial.ipynb|
\begin{figure*}[t]
	\includegraphics[width=17cm]{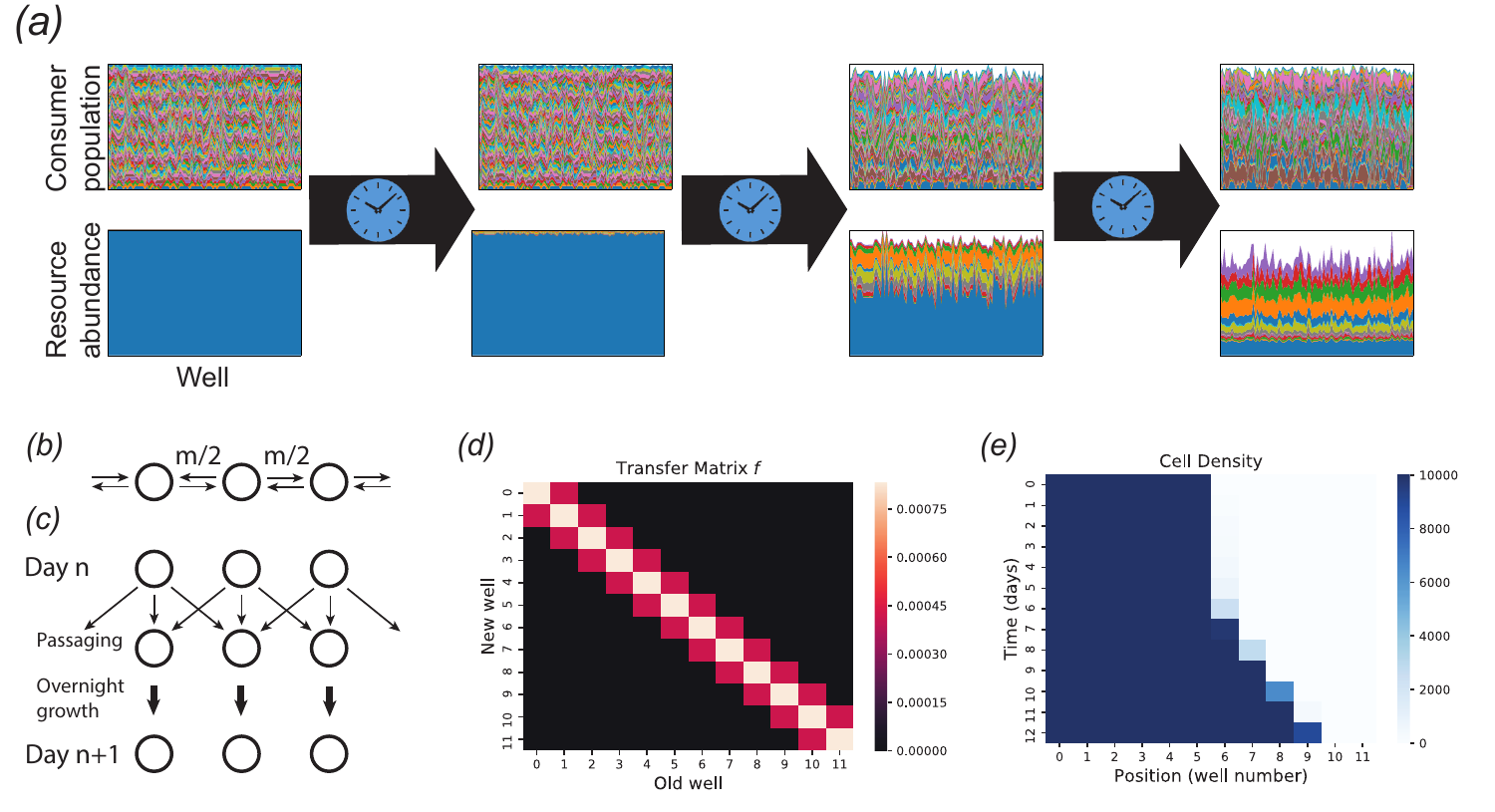}
	\caption{{\bf Propagating and passaging.} \emph{(a)} System state after successive applications of the \protect\UseVerb{prop} method to a plate with $n = 100$ wells, with a single externally supplied resource (blue). Each column of a panel represents a different well, and the height of each colored patch represents the abundance of a different consumer species or resource type. Each panel is normalized so that the sample with the largest total biomass or total resource concentration spans the entire panel. As time passes, the resources become more diverse due to the generation of metabolic byproducts, while the consumers become less diverse through competitive exclusion. \emph{(b)} Modeling spatial structure with a stepping stone model. At each time step, each cell in a given well can migrate to neighboring wells with probability $m$. \emph{(c)} Implementation of stepping stone model in a 96-well plate. Every day, the communities are passaged to fresh wells, with a fraction $f_0 (1-m)$ transferred to the corresponding position in the new set of wells, and $f_0 m$ divided equally between the two nearest neighbors, where $f_0$ is an overall dilution factor. \emph{(d)} Transfer matrix $f$ implementing the stepping stone protocol. \emph{(e)} Simulated range expansion using successive applications of the \protect\UseVerb{prop} and \protect\UseVerb{pass} methods, with the transfer matrix from the previous panel. See the Jupyter notebook \protect\UseVerb{tut} included with the package for all simulation details.}
	\label{propagate}
\end{figure*}

\subsubsection*{Consumer preferences $c_{i\alpha}$}
We choose consumer preferences $c_{i\alpha}$ as follows. As stated earlier, we assume that each specialist family has a preference for one resource class $A$ (where $A=1 \ldots  F$) with $0 \le F \le T$, and we denote the consumer coefficients for this family by $c_{i \alpha}^A$. We also consider generalists that have no preferences, with consumer coefficients $c_{i \alpha}^{\mathrm{gen}}$. The $c_{i \alpha}^A$ can be drawn from one of three probability distributions : (i) a Normal/Gaussian distribution, (ii) a Gamma distribution (which ensure positivity of the coefficients), and (iii)  a Bernoulli distribution with binary preference levels. Fig~\ref{params} shows examples of all three models.

The Gaussian model is parameterized in terms of the mean $\mu_c = \left\langle \sum_\alpha c_{i\alpha}\right\rangle$ and variance $\sigma_c^2 = {\rm var} \left( \sum_\alpha c_{i\alpha}\right)$ of the total consumption capacity, and a parameter $q$ that controls how specialized each family is for its preferred resource class. In the generalist family, the mean and variance of $c_{i\alpha}$ are the same for all resources, and are given by
\begin{align}
\< c_{i \alpha}^{\mathrm{gen}} \>&={\mu_c \over M}\\
\< (\delta c_{i\alpha}^{\rm gen})^2 \> &={\sigma_c^2 \over M}.
\end{align}
where $\delta c_{i\alpha}^{\rm gen} = c_{i\alpha}^{\rm gen} - \< c_{i \alpha}^{\mathrm{gen}} \>$ is the deviation from the mean value. The specialist families sample from a distribution with a larger mean for resources in their preferred class:
\begin{align}
\< c_{i \alpha}^A \>=
\begin{cases}
{\mu_c \over M}\left[1+\frac{M-M_A}{M_A} q\right], & \text{if}\ \alpha \in \mathbf{A} \\
{\mu_c \over M}(1-q), & \text{otherwise},
\end{cases}
\end{align}
where $M_A$ is the number of resources in class $A$ and $\mathbf{A}$ is the set of resource indices in class $A$. The variances are likewise larger for the preferred class:
\begin{align}
\< (\delta c_{i \alpha}^A)^2 \>=
\begin{cases}
{\sigma_c^2 \over M}\left[1+\frac{M-M_A}{M_A} q\right], & \text{if}\ \alpha \in \mathbf{A} \\
{\sigma_c^2 \over M}(1-q), & \text{otherwise}.
\end{cases}
\end{align}
This makes it possible to construct pure specialist families with no off-target consumption by setting $q = 1$. Note that this is different from the original version of this model in \cite{marsland2018available}, where all the variances were chosen to be identical.

We also consider the case where consumer preferences are drawn from Gamma distributions, which guarantee that all coefficients are positive. Since the Gamma distribution only has two parameters, it is fully determined once the mean and variance are specified. We parameterize the mean and variance for this model in the same way as for the Gaussian model. 

In the binary model, there are only two possible values for each $c_{i\alpha}$: a low level $\frac{c_0}{M}$ and a high level $\frac{c_0}{M} + c_1$. The elements of $c_{i \alpha}^A$ are given by
\begin{align}
c_{i \alpha}^A = \frac{c_0}{M} + c_1 X_{i \alpha},
\end{align}
where $X_{i \alpha}$ is a binary random variable that equals 1 with probability
\begin{align}
p_{i \alpha}^A =
\begin{cases}
{\mu_c \over M c_1}\left[1+\frac{M-M_A}{M_A}q\right], & \text{if}\ \alpha \in A \\
{\mu_c \over M c_1}(1-q), & \text{otherwise}
\end{cases}
\label{eq:p}
\end{align}
for the specialist families, and
\begin{align}
p_{i \alpha}^\mathrm{gen}=\frac{\mu_c}{Mc_1}
\end{align}
for the generalists. Note that the variance in each family is $\<(\delta c_{i \alpha}^A)^2\> = c_1^2 p_{i\alpha}^A (1-p_{i\alpha}^A) \sim c_1^2 p_{i\alpha}^A$ for large $M$, which depends on $q$ in the same way as the variances in the Gaussian case.

\subsubsection*{Metabolic matrix $D_{\alpha\beta}$}

We choose the metabolic matrix $D_{\alpha\beta}$ according to a three-tiered secretion model illustrated in Fig~\ref{params}. The
first tier is a preferred class of `waste' products, such as carboyxlic acids for fermentative and respiro-fermentative bacteria, with $M_w$ members. The second tier contains byproducts of the same class as the input resource. For example, this could be attributed to the partial oxidation
of sugars into sugar alcohols, or the antiporter behavior of various amino acid
transporters. The third tier includes everything else. We encode this structure in $D_{\alpha\beta}$ by 
sampling each column $\beta$ of the matrix from a Dirichlet distribution with concentration parameters
$d_{\alpha\beta}$ that depend on the byproduct tier, so that on average a fraction $f_w$ of the secreted flux goes to the first tier, while a fraction $f_s$ goes to the second tier, and the rest goes to the third. The Dirichlet distribution has the property that each sampled vector sums to 1, making it a natural way of randomly allocating a fixed total quantity (such as the total secretion flux from a given input). To write the expressions for these parameters explicitly, we let $A(\alpha)$ represent the class containing resource $\alpha$, and let $w$ represent the `waste' class. We also introduce a parameter $s$ that controls the sparsity of the reaction network, ranging from a dense network with all-to-all connection when $s \to 0$, to maximal sparsity with each input resource having just one randomly chosen output resource as $s \to 1$. With this notation, we have
\begin{align}
D_{\alpha\beta} &= {\rm Dir}(d_{1\beta},d_{2\beta},d_{3\beta},\dots,d_{M\beta})_\alpha\\
d_{\alpha\beta} &=
\begin{cases}
\frac{f_w}{sM_w}, & \text{if}\, A(\beta) \neq w \text{ and } A(\alpha) = w\\
\frac{f_s}{sM_{A(\beta)}}, & \text{if}\,  A(\beta) \neq w \text{ and } A(\alpha) = A(\beta) \\
\frac{1-f_s-f_w}{s(M-M_{A(\beta)}-M_w)}, & \text{if}\,  A(\beta) \neq w \text{ and } A(\alpha) \neq A(\beta)  \\
\frac{f_w+f_s}{sM_w}, & \text{if}\, A(\beta) = w \text{ and } A(\alpha) = w\\
\frac{1-f_w-f_s}{s(M-M_w)}, & \text{if}\, A(\beta) = w \text{ and } A(\alpha) \neq w.
\end{cases}
\end{align}
The final two lines handle the case when the `waste' type is being consumed. For these columns, the first and second tiers are identical. This led to an ambiguity in the expression presented in the Supporting Information of \cite{marsland2018available}, which we have now clarified by treating this case separately. Note that in the third line, it is implicit that $A(\alpha)\neq w$, since $A(\alpha)=w$ is covered in the first line.

\subsection*{Propagation in time}
Once an instance of the \verb+Community+ class is initialized, its state can be propagated forward in time using the \verb+Propagate+ method, as illustrated in Fig~\ref{fig1}. Since the dynamical equations and parameters were supplied at initialization, the only required argument for this method is the time $T$. When the method is invoked, the state and parameters for each well are sent to different CPU's (as many as are available) using the \verb+Pool.map+ function from the \verb+multiprocessing+ module in the Python standard library. Then the dynamical equations are integrated using the \verb+odeint+ function from SciPy, which calls the LSODA solver from the FORTRAN library ODEPACK \cite{scipy,hindmarsh1983odepack}. 

\begin{figure*}[t]
	\includegraphics[width=17cm]{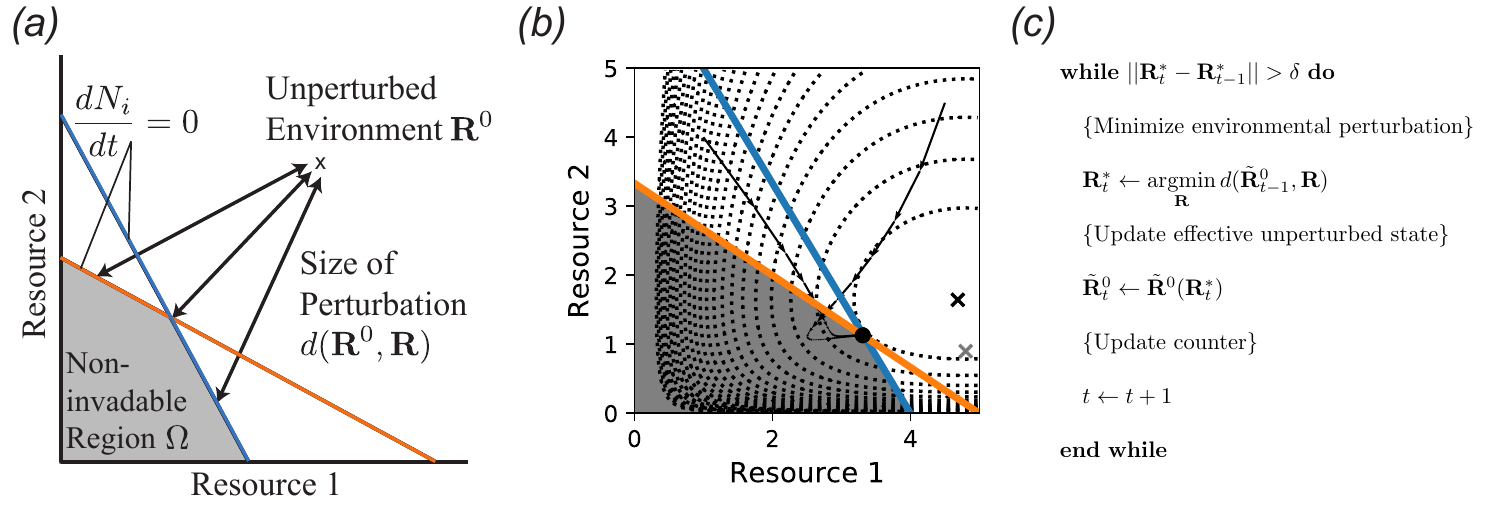}
	\caption{{\bf An expectation-maximization algorithm for finding noninvadable stationary states.} \emph{(a)} Noninvadable states by definition can only exist in the region $\Omega$ of resource space where the growth rate $dN_i/dt$ of each species $i$ is zero or negative. Here, the blue and orange lines represent the combinations of resource abundances leading to zero growth rate for two different consumer species, so the noninvadable region is the space beneath both of the lines. Within this region, a recently discovered duality implies that the stationary state $\mathbf{R}^*$ locally minimizes the dissimilarity $d(\mathbf{R}^0,\mathbf{R})$ with respect to the fixed point $\mathbf{R}^0$ of the intrinsic environmental dynamics \cite{mehta2018constrained,Marsland2019a}. \emph{(b)} Metabolic byproducts move the relevant unperturbed state from $\mathbf{R}^0$ (gray `x') to $\tilde{\mathbf{R}}^0(\mathbf{R})$ (black `x'), which is itself a function of the current environmental conditions. Dotted contour lines represent $d(\tilde{\mathbf{R}}^0(\mathbf{R}^*),\mathbf{R})$, and arrows are two trajectories of the population dynamics starting from the unperturbed environmental state with two different sets of initial consumer population sizes. See main text and Appendix for model details and parameters. \emph{(c)} Pseudocode for self-consistently computing $\mathbf{R}^*$ and $\tilde{\mathbf{R}}^0(\mathbf{R}^*)$, which is identical to standard expectation-maximization algorithms employed for problems with latent variables in machine learning.}
	\label{EM}
\end{figure*}

If the initial population size of a species equals zero, but its per-capita growth rate is positive under current environmental conditions, the limited precision of any numerical solver will enable that species to spontaneously invade the community. To prevent this from happening, the \verb+Propagate+ method comes with an option \verb+compress_species+ (set to \verb+True+ by default), which reduces the dimensionality of the state and parameters before invoking the solver by removing all references to extinct species. Compression requires deciding whether each dimension of each parameter array corresponds to species or resources. This information is built in to the package for the MicroCRM, so \verb+c+, \verb+D+,\verb+w+, \verb+g+, \verb+m+, \verb+l+, \verb+R0+, \verb+tau+ and \verb+r+ are all automatically handled correctly according to their definitions in that context. If custom dynamics are used, a dictionary of parameter dimensions must be supplied to the optional argument \verb|dimensions| of \verb+Community+ when the instance is initialized, listing which parameters are of length S, length M, or of shape SxM, SxS, or MxM, respectively.

\subsection*{Passaging to fresh plates}
The second method that can be invoked on a \verb+Community+ instance is \verb+Passage+, which simulates pipetting of cultures to fresh wells in a typical 96-well plate experiment. The only required argument for \verb+Passage+ is a two-dimensional array \verb+f+, whose elements $f_{\mu\nu}$ specify the fraction of the contents of well $\nu$ from the old plate that should be transfered to well $\mu$ on the new plate. The method also contains an option \verb+refresh_resource+, set to \verb+True+ by default, that supplies the new plate with the same initial resource concentrations as the original one. This is the most direct way of simulating actual 96-well plate experiments, where the resources are resupplied in discrete intervals at each passaging step.

This method facilitates simulation of various kinds of mixing or coalescence experiments, as well as metacommunity dynamics of weakly coupled local communities. Fig~\ref{propagate} illustrates how this feature can capture coarse-grained spatial structure, following an experimental protocol developed for mimicking range expansions in 96-well plates \cite{datta2013range}. 

In addition, passaging helps to stabilize long simulations, and simulate demographic noise, by setting all cell counts to integer values. The species abundances $N_i$ are converted to absolute cell counts through a conversion factor \verb+scale+, which is by default set to $10^6$. Then the new integer cell counts are generated by multinomial sampling based on the average number of cells $\sum_\nu f_{\mu\nu}(N_i)_\nu$ of species $i$ transferred to well $\mu$.  One source of numerical instability in ecological models is the exponentially small values reached by population sizes headed for extinction. The multinomial sampling ensures that these population sizes are fixed at zero when they become significantly smaller than 1 in absolute units. Thus a simple way of avoiding instability in a continuously resupplied chemostat model is to periodically call \verb+Passage+ with \verb+f+ set to the identity matrix and \verb+refresh_resource+ set to \verb+False+.

Since the most common experiments involve many iterations of identical \verb+Passage+ and \verb+Propagate+ steps, the package also includes a method \verb+RunExperiment(f,T,np)+, which applies a given transfer matrix \verb+f+ and propagation time \verb+T+ for \verb+np+ iterations, saving a snapshot of the plate after each propagation step. If passaging is being used simply to stabilize the integration as discussed above, and not to simulate a batch culture setting, then the value of $T$ does not affect the results (as long as it is short enough to successfully eliminate instabilities). In this case, this variable mainly serves to control the time-resolution of the resulting timeseries.

%The default value of the scale factor \verb|scale| can be changed using an optional argument when initializing a new plate. This parameter can also be adjusted using an optional argument in the \verb|Passage| method itself, as well as in \verb|RunExperiment|. The main effect of changing the scale factor is to control the amount of demographic noise introduced by the discretization.

\subsection*{Finding equilibrium points with convex optimization and expectation maximization}

\begin{figure*}[t!]
	\includegraphics[width=17cm]{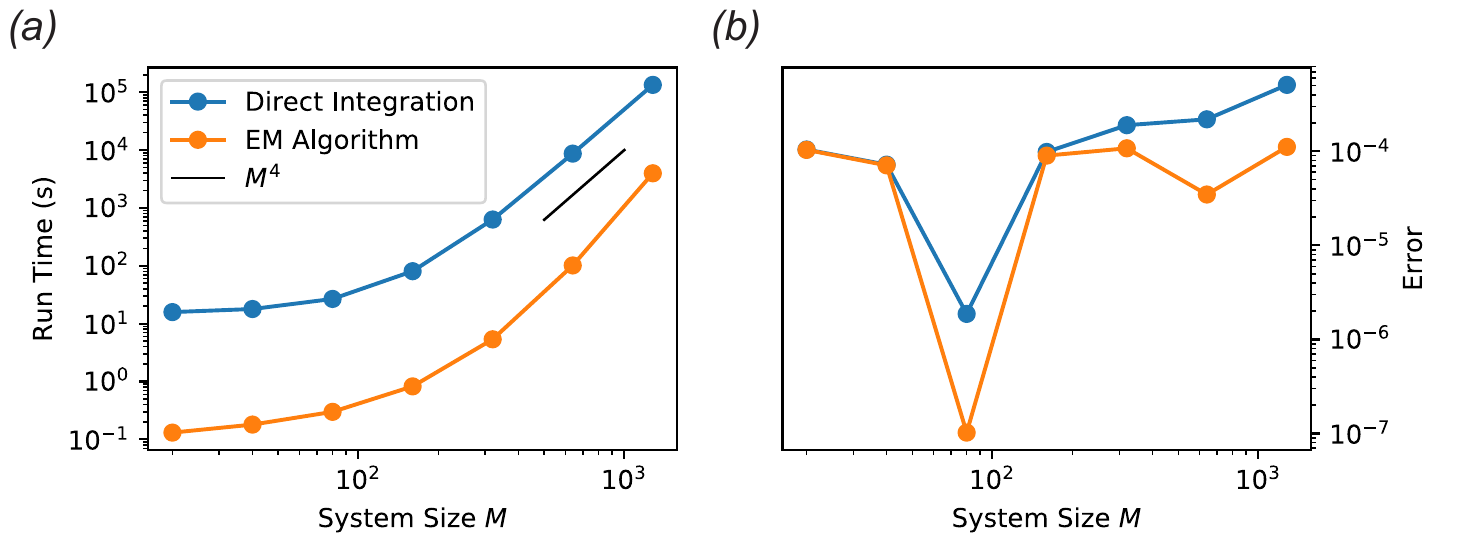}
	\caption{{\bf Performance of EM algorithm versus ODE integration.} The steady state of the MicroCRM was computed by direct ODE integration and with our new EM algorithm for a range of values of the number of resource types $M$. The initial number of species $S$ was set equal to $M$, and a single resource type was externally supplied with intrinsic fixed point $R^0_1 = 10M$ ($R^0_i = 0$ for all $i>1$). The absolute error tolerance of the integrator was set to $10^{-4}$, and the convergence tolerance for the EM algorithm was set to $\delta = 10^{-7}$. See `scripts' folder in the `EM-algorithm' branch of the GitHub repository for the rest of the parameters, which were held fixed for all simulations. \emph(a)~Total computation time for 10 realizations. \emph{(b)}~Final root-mean-square per-capita deviation of the growth rate from zero (`Error') over all surviving species in all 10 samples.}
	\label{fig:EM}
\end{figure*}

In many experimental contexts, one is interested in the stable community structure reached after a long period of constant environmental conditions. Numerical integration of the dynamical equations is an inefficient way to identify these equilibrium points. When the species diversity and number of resource types are high, the equations typically contain complex transient dynamics spanning a large range of time scales. This transient behavior is often irrelevant to the identification of the final equilibrium point, and wastes significant computation time. For a typical implementation of the MicroCRM with Type-I response and a $1,280\times 1,280$ binary consumer matrix, integrating to the steady state takes about 37 hours on standard hardware. The computation time appears to scale asymptotically as $M^4$ when the number of species $S$ and the number of resources $M$ are changed simultaneously, as shown in Fig.~\ref{fig:EM}.

To address this problem, we have developed an algorithm for identifying equilibrium points directly, without integration through the transient. This algorithm is implemented in Community Simulator as the method \verb|SteadyState|, and is illustrated in Fig.~\ref{EM}. Under the same test conditions, this algorithm converges between one and two orders of magnitude faster than numerical integration, as shown in Fig.~\ref{fig:EM}, facilitating rapid hypothesis evaluation and iteration in large ecosystems. Note that while all other features of the Community Simulator depend only on packages included with a standard Anaconda installation (\url{https://www.anaconda.com/}), \verb|SteadyState| additionally requires prior installation of a convex optimization package called CVXPY (\url{https://www.cvxpy.org/}).

The algorithm exploits a recently discovered duality between consumer resource models and constrained optimization over resource space \cite{mehta2018constrained,Marsland2019a}, which generalizes a minimization principle originally identified by MacArthur in the context of his original Consumer Resource Model \cite{MacArthur1970}. This duality applies to a wide class of consumer-resource type models, requiring only that the environmentally mediated interactions between pairs of consumer species are symmetric \cite{Marsland2019a}. 

For models in this class, it was shown that the vector of resource abundances $\mathbf{R}^*$ in every stable equilibrium state locally minimizes a measure $d(\mathbf{R}^0,\mathbf{R})$ of the dissimilarity between the current resource abundances $\mathbf{R}$ and the supply point $\mathbf{R}^0$ (defined in general by $h_\alpha(R_\alpha^0) = 0$), subject to the constraint that all consumer growth rates are zero or negative ($dN_i/dt\leq 0$ for all $i$). Instances of the MicroCRM with Type I resource consumption, no metabolic regulation and $l_\alpha = 0$ fall into this class. For externally supplied resources, $d$ turns out to be a weighted Kullback-Leibler (KL) divergence:
\begin{align}
d^{\rm external}(\mathbf{R}^0,\mathbf{R}) = \sum_{\beta} w_\beta \tau_\beta^{-1} \left[R^0_\beta \ln \frac{R^0_\beta}{R_\beta} - (R^0_\beta - R_\beta)\right]
\end{align} 
while for self-renewing resources, it is a weighted Euclidean distance \cite{Marsland2019a}:
\begin{align}
d^{\rm self-renewing}(\mathbf{R}^0,\mathbf{R}) = \sum_{\beta} w_\beta r_\beta (R^0_\beta - R_\beta)^2.
\end{align}
The equilibrium consumer populations $N_i^*$ are the Lagrange multipliers that enforce the constraints. For models with Type-I response, the non-invadable region is convex, allowing for efficient solution of the optimization problem using the Python package CVXPY \cite{cvxpy,cvxpy_rewriting}.

The duality does not strictly apply to other variants of the MicroCRM. In particular, byproduct secretion breaks the symmetry of the effective interactions between consumers whenever $l_\alpha >0$ for some resource $\alpha$ . The duality can be recovered, however, if the equilibrium point $\mathbf{R}^0$ of the intrinsic resource dynamics is changed to a new value $\tilde{\mathbf{R}}^0(\mathbf{R}^*)$, which accounts for the extra resources produced by the consumer species when the system is at its equilibrium state $\mathbf{R}^*$ \cite{Marsland2019a}. This accounting can be done in a variety of ways that all successfully recover the duality, with varying degrees of computational efficiency. The currently implemented form is
\begin{align}
\tilde{R}^0_\alpha(\mathbf{R}^*) = R^0_\alpha + \sum_{\beta\neq \alpha} \frac{Q^{-1}_{\alpha\beta}}{Q^{-1}_{\beta\beta}}(R_\beta^0 - R^*_\beta)
\end{align}
where
\begin{align}
Q_{\alpha\beta} = \delta_{\alpha\beta} - l_\beta D_{\alpha\beta} \frac{w_\beta}{w_\alpha}
\end{align}
and $Q^{-1}_{\alpha\beta}$ are the elements of the matrix inverse of $Q_{\alpha\beta}$, satisfying $\sum_{\beta} Q^{-1}_{\alpha\beta}Q_{\beta\gamma} = \delta_{\alpha\gamma}$. With these definitions, one can show that the MicroCRM with Type I consumption and no regulation minimizes the objective function
\begin{align}
d^{\rm byproducts}(\tilde{\mathbf{R}}^0,\mathbf{R}) =  \sum_{\beta} \tilde{w}_\beta \tau_\beta^{-1} \left[\tilde{R}^0_\beta \ln \frac{\tilde{R}^0_\beta}{R_\beta} - (\tilde{R}^0_\beta - R_\beta)\right]
\end{align}
where
\begin{align}
\tilde{w}_\alpha = Q^{-1}_{\alpha\alpha} (1-l_\alpha)\tau_\alpha^{-1} w_\alpha.
\end{align}

To find $\mathbf{R}^*$, one must now self-consistently solve the following equation:
\begin{align}
\label{eq:em1}
\mathbf{R}^* = \underset{\mathbf{R}}{\rm argmin}\, d(\tilde{\mathbf{R}}^0(\mathbf{R}^*),\mathbf{R}).
\end{align}
The structure of this problem is mathematically equivalent to a standard task in machine learning, where one attempts to infer model parameters from partial data \cite{mehta2018high}. These parameters $\theta$ specify a multivariate probability distribution $p(\mathbf{y}|\theta)$ for a set of measurements $\mathbf{y}$. A standard way of estimating the parameters is to compute the values $\hat{\theta}$ that maximize the likelihood of the data: $\hat{\theta} = \underset{\theta}{\rm argmax} \,p(\mathbf{y}|\theta)$. But if one actually has access to only a subset $\mathbf{x}$ of the measurement results, then the values $\mathbf{z}$ of the remaining quantities must also be estimated in order to perform this optimization. Ideally, one would use the statistical model with the optimal parameters $\hat{\theta}$ for this task. In the simplest case, where the value of $\mathbf{z}$ can be inferred with certainty given $\theta$ and $\mathbf{x}$, this results in the following self-consistency equation:
\begin{align}
\label{eq:em2}
\hat{\theta} = \underset{\theta}{\rm argmax}\, p(\mathbf{x},\mathbf{z}(\hat{\theta})|\theta).
\end{align}
This is identical in form to Eq.~\ref{eq:em1}, where the parameters $\theta$ become the resource concentrations $\mathbf{R}$ and the estimated latent variables $\mathbf{z}$ become the effective unperturbed state $\tilde{\mathbf{R}}^0$. The observed data $\mathbf{x}$ become the model parameters, which are implicitly used in the calculation of $d$ and $\tilde{\mathbf{R}}$. We can think of the environmental perturbation $d$ as a statistical potential or ``free energy'' $-\ln p$, which is minimized when $p$ is maximized.

Eq.~\ref{eq:em2} can be solved by a standard iterative approach called Expectation Maximization \cite{mehta2018high}. At each iteration $t$, the latent variable $\mathbf{z}_t$ is computed from the previous estimate $\hat{\theta}_{t-1}$ of $\hat{\theta}$, and then the new parameter estimate $\hat{\theta}_t$ is found by maximizing $p(\mathbf{x},\mathbf{z}_t|\theta)$. Fig.~\ref{EM}\emph{(c)} contains pseudocode for this algorithm as applied to our ecological problem, which was also reported previously in \cite{Marsland2019a}. 

This algorithm fails to converge at low resource supply levels, because both arguments must be positive when $d$ is a weighted KL divergence, but $\tilde{R}^0_t$ can temporarily become negative under these conditions. To solve this issue, we replaced update step for $\tilde{R}^0_t$ by
\begin{algorithmic}
	\centering
	\STATE $\tilde{\mathbf{R}}^0_t \gets \alpha\tilde{\mathbf{R}}^0(\mathbf{R}^*_t) + (1-\alpha) \tilde{\mathbf{R}}^0_{t-1}$ 
\end{algorithmic}
where $\alpha$ is a constant rate, equivalent to the ``learning rate'' in machine learning \cite{mehta2018high}.

Default values of the tolerance $\delta$ and learning rate $\alpha$ are set to $10^{-7}$ and 0.5, respectively, which give robust convergence for typical simulation scenarios. They can be adjusted as optional arguments of the \verb|SteadyState| method. 

This algorithm can be applied to any consumer-resource type model, including models beyond the MicroCRM framework, with non-substitutable resources \cite{Marsland2019a}. But the enhanced efficiency of the new approach requires that the optimization problem be convex. In the Community Simulator package, the algorithm is only implemented for Type-I response with no metabolic regulation, where convexity is guaranteed. For more complex models, the differential equations must be numerically integrated using the \verb|Propagate| method discussed above. 

In the tutorial notebook included with the package, we show that the MicroCRM can be bistable if the externally supplied resources are insufficient to directly support growth of any consumer species. In this scenario the state with all consumers extinct is a stable equilibrium of the dynamics, and another stable equilibrium with persisting consumers is also possible that relies on the metabolic byproducts. In this scenario \verb|SteadyState| method can find either of the two equilibria, depending on the initial estimate of $\tilde{\mathbf{R}}^0$, which can be set by an optional argument. If this initial condition is sufficiently close to the actual equilibrium state $\mathbf{R}^0$ of the intrinsic resource dynamics, the method ends in the state with the consumers extinct, where $\tilde{\mathbf{R}}^0 = \mathbf{R}^0 = \mathbf{R}^*$. But if the initial condition is not deliberately tuned to be close to $\mathbf{R}^0$, we find that the method typically finds the other state where some consumers survive. 
%\emph{(d)}  Convergence dynamics of the effective unperturbed state $\tilde{\mathbf{R}}^0_t$ and the corresponding equilibrium state $\mathbf{R}^*_t$ to their true values (dotted lines) for four arbitrarily chosen resource types (out of $M=100$ resource types present in the simulation). Initial values of all elements of $\tilde{\mathbf{R}}^0$ were arbitrarily set equal to 10. \emph{(e)} The steady state of the MicroCRM was computed by direct ODE integration and with our new EM algorithm for a range of values of the number of resource types $M$. The initial number of species $S$ was set equal to $M$, and a single resource type was externally supplied with intrinsic fixed point $R^0_1 = 10M$ ($R^0_i = 0$ for all $i>1$). See main text and Appendix for implementation details and for the rest of the parameters, which were held fixed for all simulations. Top: Total computation time for 10 realizations. Bottom: Final root-mean-square per-capita deviation of the growth rate from zero (`Error') over all surviving species in all 10 samples. \emph{(f)} The EM algorithm was executed for different values of the unperturbed level $R^0_1$ of the externally supplied resource, at fixed system size $M = 100$. Top: Total computation time for 100 independent realizations. Bottom: Fraction of realizations for which algorithm failed to converge.

\section*{Discussion}
One interesting future direction to explore is integrating the Community Simulator with methods for directly analyzing  Microbiome sequencing data. For example, there has been a renewed interest in statistical techniques such as Approximate Bayesian Computation (ABC) for understanding ecology and evolution \cite{csillery2010approximate}. In ABC, the need to exactly calculate complicated likelihood functions -- often a prerequisite for many statistical techniques -- is replaced with the calculation of summary statistics and numerical simulations. For this reason, the Community Simulator Python package is ideally suited to form the backbone of new inference techniques for trying to related  ecological processes to observed abundance patterns in microbial ecosystems. 

\section*{Conclusion}
We hope that the Community Simulator will become a valuable resource for the microbial ecology community. It has already played an important role in our own work. The package initially facilitated the systematic evaluation of the robustness of results to different modeling assumptions in a study of the effects of total energy influx on community structure, diversity and function \cite{Marsland2019a}. More recently, the convex optimization approach has made it possible to perform more than 100,000 independent simulations in a reinterpretation and extension of Robert May's classic work on diversity and stability \cite{Cui2019,May1972}. We have also employed the package to reproduce large-scale patterns in microbial biodiversity from the Human Microbiome Project, Earth Microbiome Project, and similar surveys \cite{marsland2019minimal}. Finally, the random matrix approach implemented in this package is amenable to analytic calculation in the limit of large numbers of species and resources, using cavity methods from the physics of disordered systems \cite{Advani2018,cui2019effect}. It is our belief that the Community Simulator will facilitate the further development of these mathematical techniques through efficient testing of new conjectures.

\section*{Availability and Requirements}
\begin{itemize}
	\item {\bf Project name}: Community Simulator
	\item {\bf Project home page}: \url{https://github.com/Emergent-Behaviors-in-Biology/community-simulator}
	\item {\bf Operating system(s)}: Linux or Mac preferred. Parallelization scheme is currently incompatible with Windows, and must be deactivated (set \verb|parallel=False| when initializing a plate) for the code to run.
	\item {\bf Programming language}: Python 3
	\item {\bf Other requirements}: Numpy 1.15+, Pandas 0.23.0+, Matplotlib 2.2.3+, SciPy 1.1.0+. \verb|SteadyState| method additionally requires CVXPY 1.0+.
	\item {\bf License}: MIT
	\item {\bf Any restrictions to use by non-academics:} None
\end{itemize}

\section*{Acknowledgements}
This work was supported by NIH NIGMS grant 1R35GM119461 and Simons Investigator in the Mathematical Modeling of Living Systems (MMLS) award to PM. We are grateful to Kirill Korolev,  Alvaro Sanchez, and Daniel Segr\`e for many useful conversations, and to Matti Gralka for testing cross-platform compatibility of the package. The performance evaluation reported in Fig~\ref{fig:EM} was performed on the Shared Computing Cluster which is administered by Boston University Research Computing Services.

% Either type in your references using
% \begin{thebibliography}{}
% Text
% \end{thebibliography}
%
% or
%
% Compile your BiBTeX database using our plos2015.bst
% style file and paste the contents of your .bbl file
% here. See http://journals.plos.org/plosone/s/latex for 

\end{document}